\documentstyle[aps,prb,multicol,epsf]{revtex}

\begin{document}
\draft
\title{Low field negative magnetoresistance in double layer structures}
\author{G.\ M.\ Minkov, A.\ V.\ Germanenko, O.\ E.\ Rut}
\address{Institute of Physics and Applied Mathematics, Ural State University,
620083 Ekaterinburg, Russia}
\author{O.\ I.\ Khrykin, V.\ I.\
Shashkin, V.\ M.\ Danil'tsev}
\address{Institute of Physics of Microstructures of RSA, 603600 N.
Novgorod, Russia}

\date{\today}
\maketitle

 \widetext
\begin{abstract}The weak localization correction to the conductivity in coupled
double layer structures is studied both experimentally and theoretically.
Statistics of closed paths has been obtained from the analysis of magnetic
field and temperature dependencies of negative magnetoresistance for magnetic
field perpendicular and parallel to the structure plane. The comparison of
experimental data with results of computer simulation of carrier motion  over
two 2D layers with scattering shows that inter-layers transitions play
decisive role in the weak localization.
\end{abstract}

\begin{multicols}{2}
\narrowtext

\section{Introduction}
\label{sec:intr} Transitions between 2D layers is one of
fundamental features of double layer structures. It changes the
quantum corrections to the conductivity, especially in a magnetic
field parallel to the layers.

It is well known \cite{Aronov} that the interference of electron
waves scattered along closed trajectories in opposite directions
(time-reversed paths) produces a negative quantum correction to
the conductivity. An external magnetic field ($\bf B$) gives the
phase difference between pairs of time-reversed paths
$\varphi=2\pi({\bf B S})/\Phi_0$, where $\Phi_0$ is the quantum
of magnetic flux, $\bf S$ is the area enclosed, and thus destroys
the interference and results in negative magnetoresistance.

In case of a single 2D layer the influence of a magnetic field is
strongly anisotropic because all the paths lie in one plane. The
magnetoresistance is maximal for $\bf B\parallel n$, where ${\bf
n}$ is the normal to 2D layer. When a magnetic field lies in the
2D layer plane, $\bf B\perp n $, the scalar product $({\bf BS})$
is zero, i.e. the magnetic field does not destroy the
interference, and the negative magnetoresistance is absent in
this magnetic field orientation. \cite{Falko}

In coupled double layer structures, the tunneling between layers
gives rise to the closed paths where an electron moves initially
over one layer then over another one and returns to the first
layer. For this paths the product (${\bf BS}$) is non-zero for
any magnetic field orientation and hence the negative
magnetoresistance  has to appear for  $\bf B\perp n $ as well.

The magnetic field dependence of the negative magnetoresistance
is determined by the statistics of closed paths, namely, by the
area distribution function, $W(\bf S)$, and area dependence of
the average length of closed paths, $\overline{L}(\bf S)$.
\cite{Pogosov,our1} Just these  statistic dependencies have been
studied in single 2D layer structures by analysis of negative
magnetoresistance measured at $\bf B\parallel n$.\cite{our1,our2}

The role of inter-layers transitions in weak localization  and
negative magnetoresistance for $\bf B\parallel n $ for multilayer
structures (superlattices) was discussed in Ref.\
\onlinecite{multi}. The closely related problem concerning the
role of inter-subbands transitions in quasi-two dimensional
structures with several subbands occupied was theoretically
studied in Ref.\ \onlinecite{intersub}.

In this paper we present the results of investigations of the
negative magnetoresistance in double layer GaAs/InGaAs structure
for different magnetic field orientations. We obtain the area
distribution functions and area dependencies of the average
lengths of the closed paths using the approach developed in
Refs.\ \onlinecite{our1,our2}. These functions are compared with
those obtained from the computer simulation of carrier motion
when inter-layers transitions are accounted for. Close agreement
shows that just the inter-layers transitions determine the
negative magnetoresistance in coupled double layer structures.

\section{experimental results}
\label{sec:exp}

\begin{figure}
\epsfclipon
 \epsfxsize=\linewidth
 \epsfbox{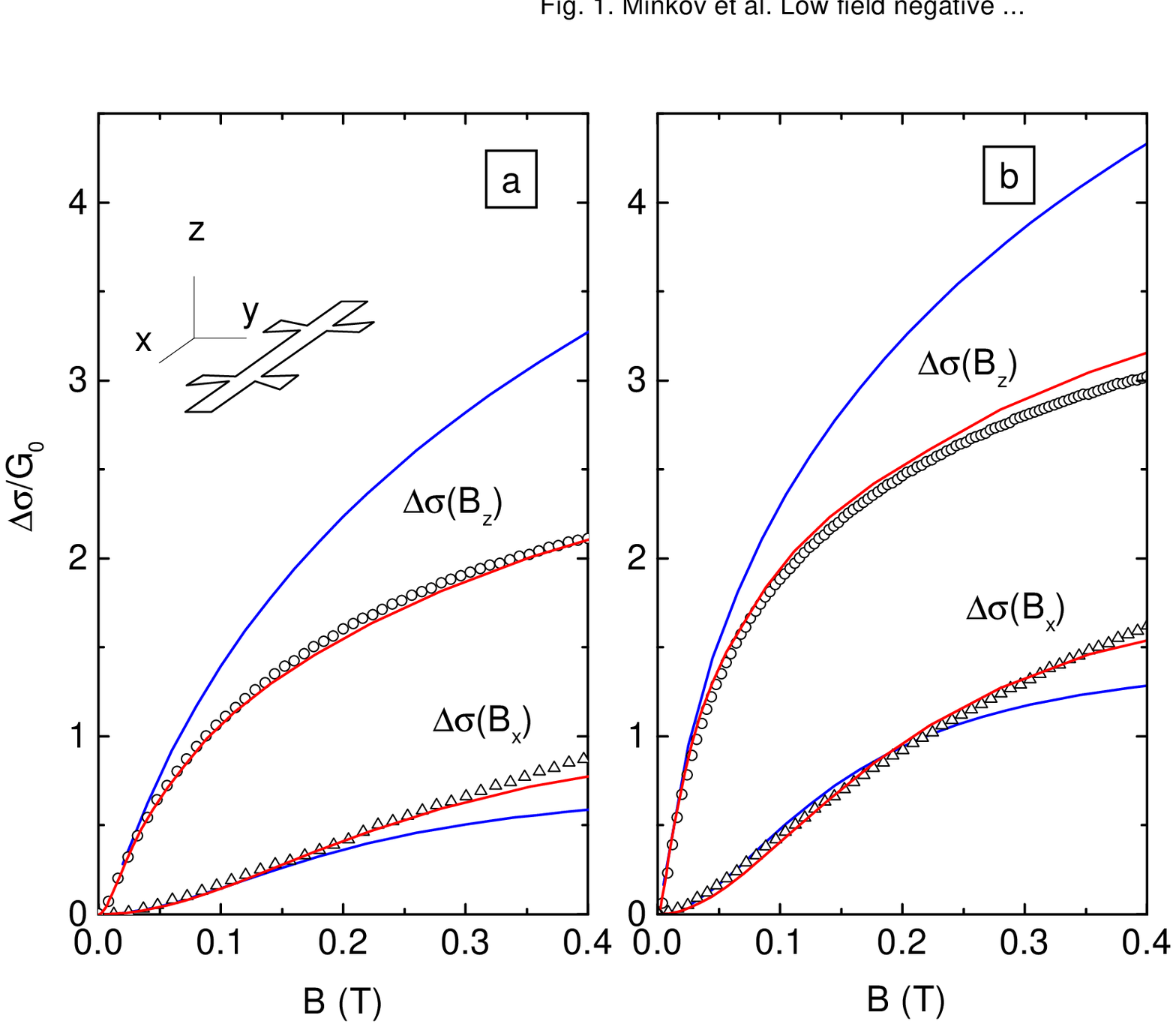}
\caption{Magnetic field dependencies of $\Delta\sigma/G_0$ for
different magnetic field orientations for $T=4.2$ (a), $1.5$ K
(b). Symbols are the experimental data, red curves are the
simulation results. Blue curves are the results of calculations
carried out according to Ref.\ \protect\onlinecite{raichev}.
Insert in (a) shows a system of coordinates.}
  \label{fig10}
\end{figure}

The double well heterostructure GaAs/In$_{x}$Ga$_{1-x}$As
 was grown by Metal-Organic Vapor Phase Epitaxy (MOVPE)
on semi-insulator GaAs substrate. The heterostructure consists of
a 0.5 $\mu$m-thick undoped GaAs epilayer, a Si $\delta$-layer, a
75~\AA\  spacer of undoped GaAs,  a 100~\AA\
In$_{0.08}$Ga$_{0.92}$As well, a 100~\AA\ barrier of undoped
GaAs, a 100~\AA\ In$_{0.08}$Ga$_{0.92}$As well, a 75~\AA\  spacer
of undoped GaAs,  a Si $\delta$-layer  and  1000~\AA\ cap layer
of undoped GaAs. The samples were {\bf mesa etched} (cut??) into
standard Hall bridges. The measurements were performed in the
temperature range $1.5 - 4.2$ K at low magnetic field up to $0.4$
T with discrete $10^{-4}$ T for two orientations: the magnetic
field was perpendicular (${\bf B}\parallel{\bf z}$) and  parallel
(${\bf B}\parallel{\bf x}$) to the structure plane (see insert in
Fig.\ \ref{fig10}). Additional high field measurements were also
made to characterize the structure. It has been found that in the
structure investigated the conductivity is determined by the
electrons in the wells. Their densities have been determined from
the Fourier analysis of the Shubnikov-de Haas oscillations and
consist of $4.5\times 10^{11}$ cm$^{-2}$ and $5.5\times
10^{11}$~cm$ ^{-2}$ in different wells. The Hall mobility was
about $\mu\simeq 4200$ cm$^2/$ (V~$\times $ sec).

The magnetic field dependencies of in-plane  magnetoconductance
\begin{equation}
 \label{sigma}
 \Delta \sigma(B)=\sigma(B)-\sigma(0)=1/\rho(B)-1/\rho(0)
\end{equation}
at magnetic field perpendicular ($\Delta \sigma(B_z)$) and
parallel ($\Delta\sigma(B_x)$) to the structure plane are
presented in Fig.\ \ref{fig10}. One can see that the negative
magnetoresistance is observed for both magnetic field
orientations and, in contrast to the case of single layer
structures, the effects are comparable in magnitude. Analysis of
behaviour of the conductivity in a wide range of temperatures
($1.5<T<20$~K) and magnetic fields ($B<6$~T) shows that at
$B<0.4-0.5$~T  the main contribution to the negative
magnetoresistance comes from the interference correction. In this
case  the magnetic field dependence of the negative
magnetoresistance is determined by the statistics of the closed
paths.\cite{Pogosov,our1}

Let us apply the method proposed in Refs.\ \onlinecite{our1,our2}
to analysis of negative magnetoresistance in the double layer
structure. Using formalism presented in Section II of Ref.\
\onlinecite{our1} one can write  the expression for  conductivity
of double layer structure with identical layers for two magnetic
field orientations as follows
\begin{eqnarray}\label{eqSbL}
  \sigma( B_i)&=& \sigma_0+\delta\sigma( B_i) \nonumber \\
  &=&\sigma_0-4\pi l^2 G_0 \int_{-\infty}^{\infty} d S_i\
 \biggl\{ W(S_i) \nonumber \\
 && \exp\left(-\frac{\overline{L}(S_i)}{l_\varphi}\right)\
 \cos\left(\frac{2\pi  B_i S_i}{\Phi_0}\right)\biggr\} ,\; i=x,\;z.
\end{eqnarray}
Here, $G_0=e^2/(2\pi^2 \hbar)$, $\sigma_0$ is the classical Drude
conductivity, $l=v_F\tau$, $l_\varphi=v_F \tau_\varphi$, $v_F$ is
the Fermi velocity, $\tau$ and  $\tau_\varphi$ are the momentum
relaxation and phase breaking time, respectively. The value of
$\overline {L}$ is the function not only of $S$ but $l_\varphi$
as well.  It was defined in Ref.\ \onlinecite{our1} by Eq.(6). It
should be mentioned that Eq.\ (\ref{eqSbL}) is valid at low
enough probability of inter-layers transitions.

Thus, for the magnetic field perpendicular to the structure plane
(${\bf B}=(0,0,B_z)$) the magnetoresistance is determined by
$z$-component of ${\bf S}$ only and for parallel magnetic field
(${\bf B}=(B_x,0,0)$) it is determined by $x$-component of ${\bf
S}$.

One can see from Eq.\ (\ref{eqSbL}) that the Fourier transform of
$\delta\sigma(B)/G_0$
\begin{eqnarray}
\Phi (S_i,l_\varphi)&=&\frac{1}{\Phi_0} \int_{-\infty}^{\infty}d
B_i\ \frac{\delta\sigma(B_i)}{G_0}
 \cos\left(\frac{2\pi B_i S_i}{\Phi_0}\right)= \nonumber \\
&=& 4\pi l^{2} W(S_i)\exp \left( -\frac{
\overline{L}(S_i)}{l_{\varphi } } \right), \label{eq6}
\end{eqnarray}
carriers an information on $W(S_i)$ and $\overline{L}(S_i)$.
Because the value of $l_\varphi$ tends to infinity when
temperature tends to zero, \cite{saturation} the extrapolation of
$\Phi (S_i,l_\varphi)$-vs-$T$ curve to $T=0$ gives the value of
$4\pi l^2  W(S_i)$. The  ratio $\overline{L} (S_i)/l_\varphi$ for
given $S_i$ can be then obtained as  $\ln( 4\pi
l^{2}W(S_i))-\ln(\Phi(S_i,l_\varphi))$.
\begin{figure}
\epsfclipon
 \epsfxsize=\linewidth
 \epsfbox{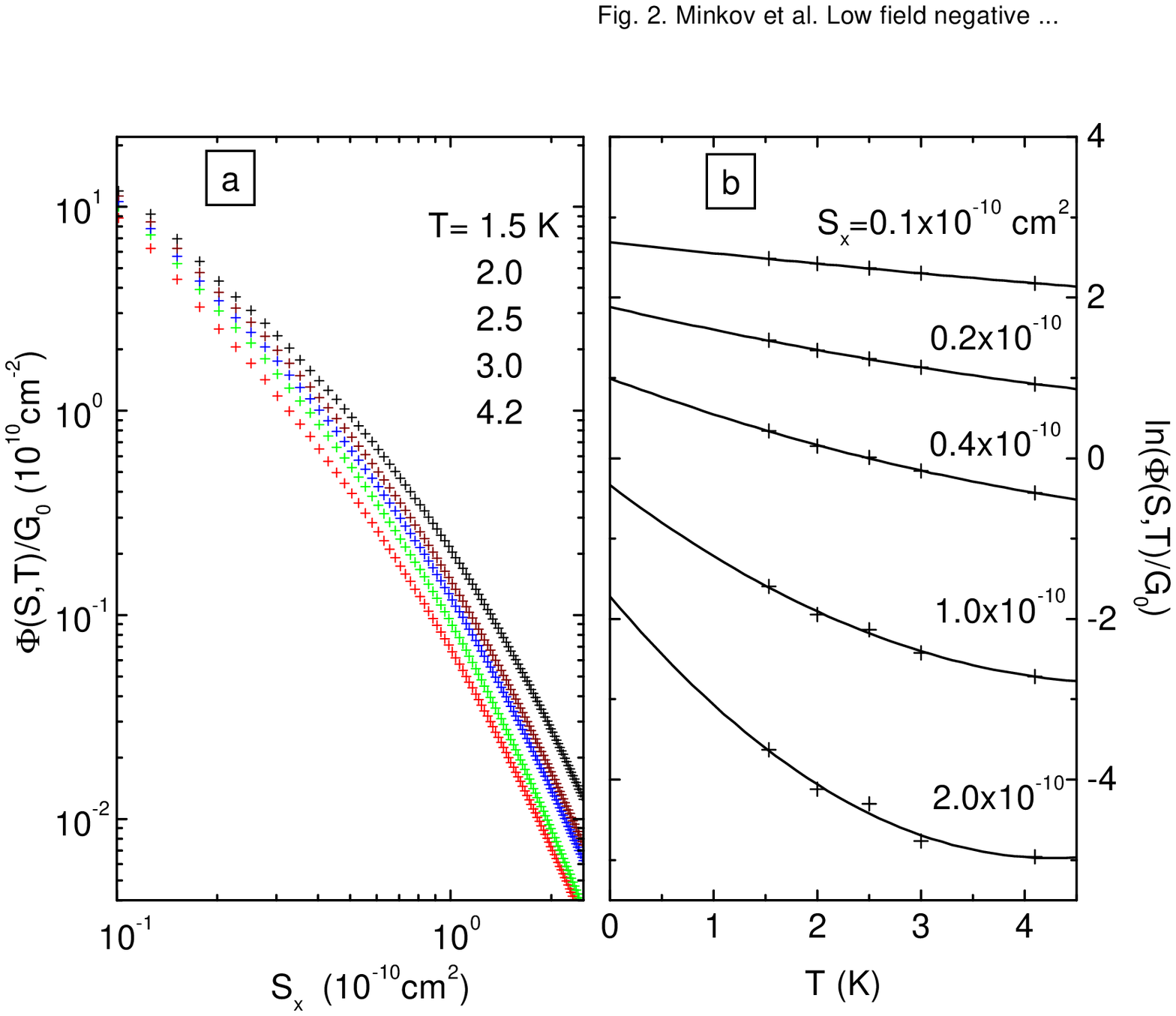}
\caption{{\bf} Area (a) and temperature (b) dependencies of the
Fourier transforms of $\delta\sigma(B_x)$. Curves in (b) show the
extrapolation of $\Phi(S,T)$ to $T=0$.}
  \label{fig20}
\end{figure}

The value  $\Delta \sigma(B)= \sigma(B)-\sigma(0)$, not
$\delta\sigma(B)$, is experimentally measured. It is clear from
Eqs. (\ref{sigma}) and (\ref{eqSbL}) that $\delta
\sigma(B)=\sigma(0)-\sigma_{0}- \Delta \sigma(B)$. To obtain
$\delta \sigma(B)$, we assume that the Drude conductivity
$\sigma_{0}$ is equal to the conductivity at T=20 K, when the
quantum corrections are small. Notice that the final results are
not sensitive to the value of  $\sigma_{0}$ practically.
Obtaining of the distribution function $W(S_x)$ from the
experimental $\delta \sigma(B_z)$ dependencies for the structure
investigated is illustrated by Fig.\ \ref{fig20}. In left panel
the Fourier transforms of $\delta \sigma(B_x)$ measured at
different temperatures are presented. Right panel shows how the
$\Phi$-vs-$T$ data have been extrapolated to $T=0$. The area
distribution function $W(S_z)$ has been obtained from the
analysis of $\delta \sigma(B_z)$ curves in a similar way.

\begin{figure}
\epsfclipon
 \epsfxsize=\linewidth
 \epsfbox{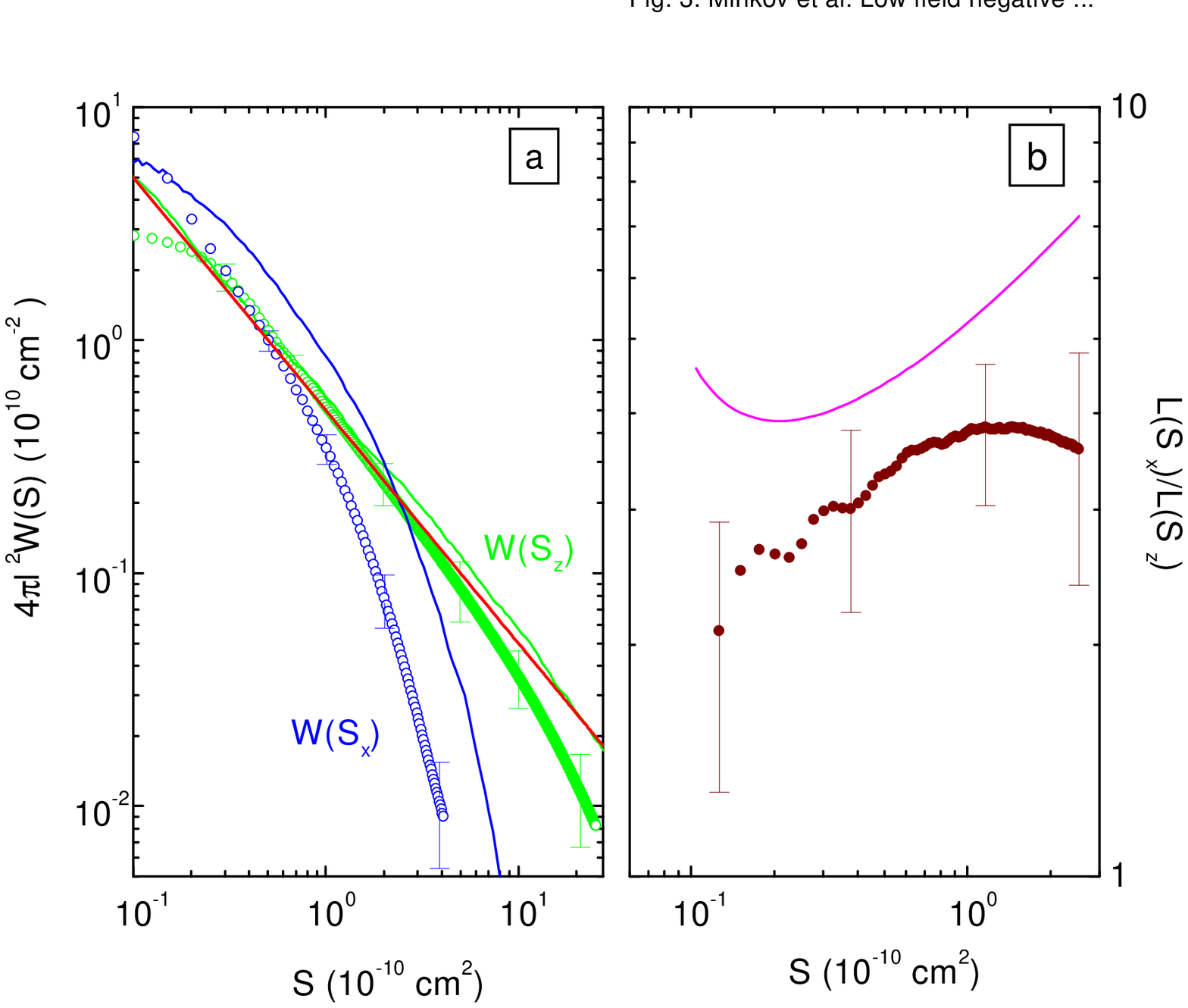}
\caption{The area distribution functions of closed paths (a) and
the area dependence of the $\overline{L}(S_x)/\overline{L}(S_z)$
ratio at $T=1.5$ K (b). The symbols are the experimental data,
curves are the  results of simulation with $t=0.1$, red line in
(a) is $(2S)^{-1}$-dependence.}
  \label{fig30}
\end{figure}

The results of data processing described above are presented in
Fig.\ \ref{fig30}a. As is seen the $4\pi l^{2}W(S_z)$ dependence
is close to $(2 S_z)^{-1}$ for $S \simeq (0.3-5) \times
10^{-10}$~cm$^{2}$. The analogous behaviour of the area
distribution function  was obtained for single 2D layer in Ref.\
\onlinecite{our2}.  The behaviour of $W(S_x)$ significantly
differs from that of $W(S_z)$. In particular,  the $W(S_x)$ curve
shows a much steeper decline for $S > 0.8\times
10^{-10}$~cm$^{2}$ . Other feature of the statistics  of the
closed paths in double layer structure is the fact that for given
$S$ the values of $\overline{L}(S_x)$ are significantly larger
than $\overline{L}(S_z)$ (Fig.\ \ref{fig30}b).

Qualitatively these peculiarities of the statistics of closed
paths in double layer structures can be understood if one
considers how trajectories with large enough length, $L\gg l/t$,
look. They are isotropically  smeared over $xy$-plane for the
distance $\sim \sqrt{L l}$, their extended area in this plane is
$s_z \sim L l$. In $yz$-plane they have size $\sim \sqrt{L l}$ in
$y$-direction and $Z_0$ (where $Z_0$ is inter-layer distance) in
$z$-direction. So, the extended area in $yz$-plane is $s_x\sim
Z_0\sqrt{L l}$. Thus, closed trajectories have significantly
larger $s_z$ than $s_x$, and the $s_z/ s_x$ ratio increases with
increasing $s$. It is clear that the behaviour of enclosed areas
$S_z$, $S_x$ is analogous. Therefore, for $S_x=S_z$ the
inequality ${W}(S_z)>{W}(S_x)$ is valid. The average length of
the trajectories $\overline{L}(S_x)$ therewith is greater than
$\overline{L}(S_z)$.

As was shown in Ref.\ \onlinecite{our1}  the distribution
function of closed paths, the area dependence of average length
of closed paths and weak localization magnetoresistance can be
obtained by computer simulation of a carrier motion over 2D
plane.

\section{Computer simulation}
\label{sec:sim}

The model double layer system is conceived as two identical plains
with randomly distributed scattering centers with a given total
cross-section. Every plane is represented as a lattice $M\times M$
with lattice parameter $a$. The scatterers are placed in a part of
the lattice sites with the use of a random number generator. We
assume that a particle moves with a constant velocity along
straight lines which happen to be terminated by collisions with
the scatterers. After every collision the particle has two
possibilities:  it passes from one plane to another with a
probability $t$ and moves over the second plane or  it remains in
the plane with probability $(1-t)$, changing the motion direction
only. If the trajectory of the particle passes near the start
point at the distance less than $d/2$ (where $d$ is a prescribed
value, which is small enough), it is perceived as being closed.
The projections of enclosed algebraic area is calculated according
to
\begin{eqnarray}\label{eqS0}
S_z=\sum_{j=1}^{N-1}\frac{y_{j+1}+y_j}{2} (x_{j+1}&-&x_j)+
\nonumber
\\ +\frac{y_{N}+y_1}{2}(x_{N}&-&x_1),
\end{eqnarray}
\begin{eqnarray}\label{eqS1}
S_x=\sum_{j=1}^{N-1}\frac{y_{j+1}+y_j}{2} (z_{j+1}&-&z_j)+
\nonumber
\\ +\frac{y_{N}+y_1}{2}(z_{N}&-&z_1),
\end{eqnarray}
where $N$ is number of collisions for given trajectory, $x_j$,
$y_j$, $z_j$ stand for coordinates of $j$-th collision, $z_j$
takes the value $0$ or $Z_0$. Otherwise the simulation details are
analogous to them described in Ref.\ \onlinecite{our1} for the
case of single 2D layer system.

All the results presented here have been obtained using the
parameters: lattice dimension is $6800\times 6800$; the number of
starts, $I_s$, is $10^6$; the total number of scatterers is about
$1.6\times 10^5$; the scattering cross section is $7$; $d=1$;
$Z_0=18$. The value of mean free path computed for such a system
is about $43\times a$. If we suppose the value of $a$ equal to
$11$ \AA, this model double layer system corresponds to the
heterostructure investigated. Namely, the mean free path is equal
to the value of $l\simeq 480$ \AA, and the value of $Z_0$ is
close to the distance between the centers of the quantum wells,
$200$ \AA.

\begin{figure}
\epsfclipon
 \epsfxsize=\linewidth
 \epsfbox{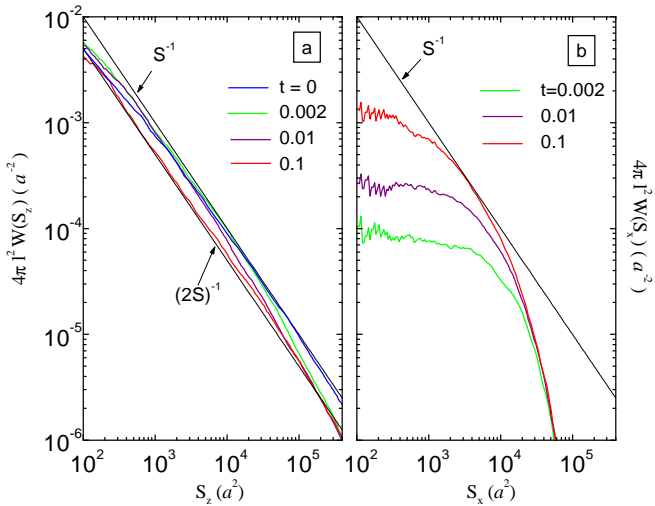}
\caption{Area distribution functions $W(S_z)$ (a) and $W(S_x)$
(b) as they have been obtained from the simulation procedure with
different $t$ values.}
  \label{fig40}
\end{figure}

The area distribution functions obtained as the result of
simulation with different inter-layers transition probabilities
are presented in Fig.\ \ref{fig40}. Let us discuss at first the
behaviour of $W(S_z)$ (Fig.\ \ref{fig40}a). For $t=0$, the $4\pi
l^2 W(S_z)$ curve corresponds to the area distribution function
for single layer. For large $S$ this curve goes close to the
$S^{-1}$-dependence, which corresponds to the ideal 2D system in
the diffusion regime. \cite{our1} The deviation, which is evident
for $S_z<10^3\:a^2$, is just due to the transition to the
ballistic regime. It is obviously that for sufficiently large
values of $t$, the probability of return to the start point has
to be twice smaller than that for $t=0$. As is seen from Fig.\
\ref{fig40}a even the value $t=0.1$ is large enough in this
sense: the corresponding curve is close to the
$(2S)^{-1}$-dependence practically in whole area range. This is
because the long trajectories with large number of passes between
layers give significant contribution to $W(S_z)$ starting from
small areas, $S_z>0.1\: l^2$. For the intermediate value of $t$
($t=0.002,\; 0.01$) the area distribution function is close to
the $S^{-1}$ function for small areas and tends to the
$(2S)^{-1}$ dependence for large ones.

The behaviour of $W(S_x)$ contrasts with that of $W(S_z)$ (Fig.\
\ref{fig40}b). At small $S_x$, $W(S_x)$ depends only weekly on
$S_x$, whereas  at large $S_x$ it decreases sharply when $S_x$
increases. Sensitivity of $W(S_x)$ to inter-layers transition
probability depends on $S_x$ value. For small $S_x$ values, when
the area distribution function is mainly determined by short
closed paths with small number of inter-layers transitions, the
value of $W(S_x)$ considerably increases with increasing $t$. For
large $S_x$, i.e. for paths with large number of inter-layer
transitions, $W(S_x)$ weakly depends on the transition
probability.

\begin{figure}
\epsfclipon
 \epsfxsize=\linewidth
 \epsfbox{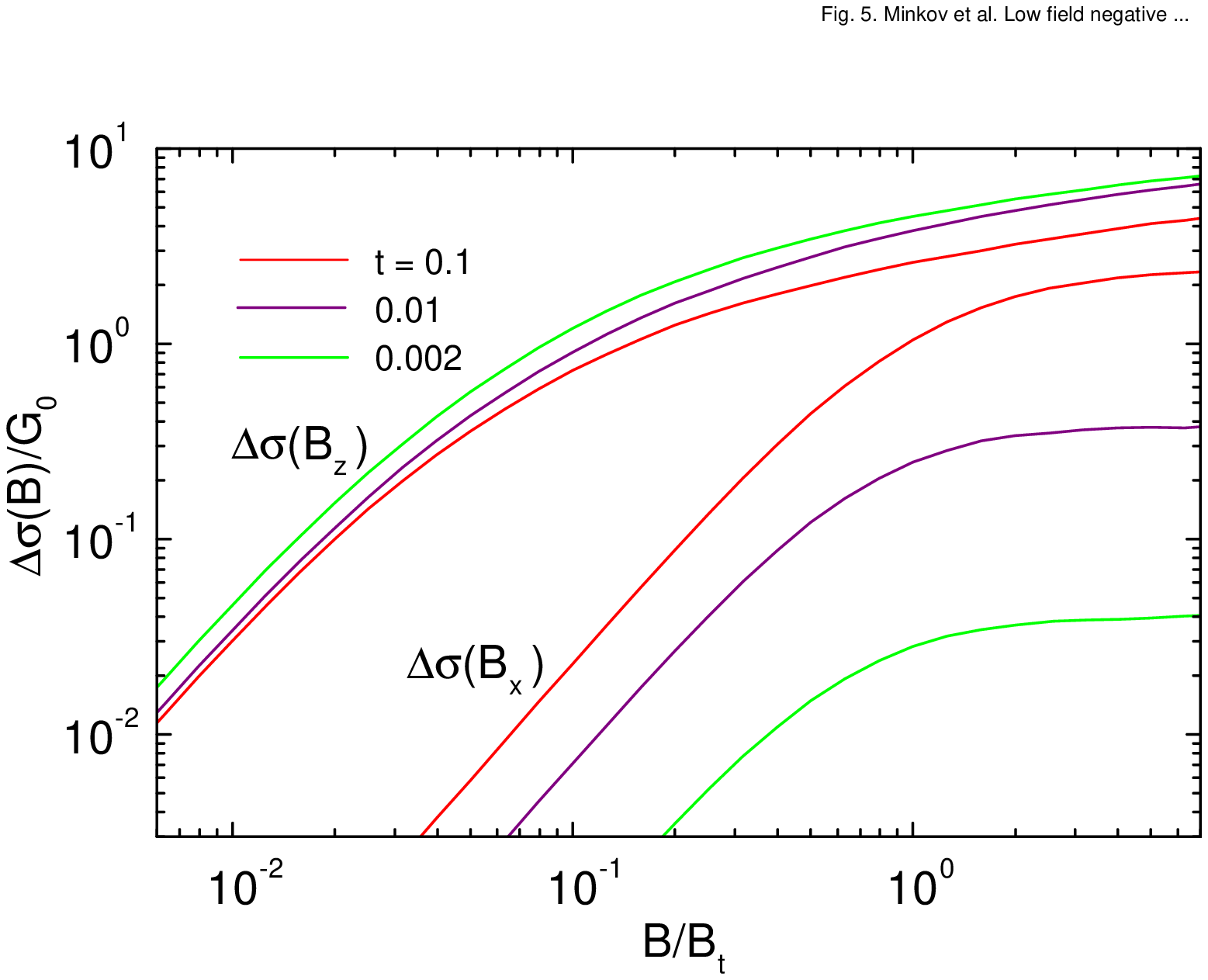}
\caption{Calculated magnetic field dependencies of $\Delta\sigma$
for different inter-layers transition probability,
$l/l\varphi=0.01$.}
  \label{fig50}
\end{figure}

Let us demonstrate how the magnetoresistance of our model 2D
system changes with changing of the inter-layers transition
probability. The theoretical $\delta\sigma (B)$ dependencies have
been calculated by summing over the contributions of every closed
path to the conductivity  in accordance with following expression
\cite{our1}
\begin{equation}
 \label{eq:s(b)}
 \frac{\delta{\sigma}(B_i)}{G_0}=\frac{2\pi l}{I_s d}
 \sum_i\cos
 \left(
 \frac{2\pi B_i S_i}{\Phi_0}
 \right)
 \exp\left(-\frac{l_i}{l_\varphi}
 \right),
\end{equation}
where $l_i$ is the length of $i$-th closed path. The results of
calculation are presented in Fig.\ \ref{fig50}, where $\Delta
\sigma (B_i)=\delta \sigma (B_i)-\delta \sigma (0)$ is plotted
against $B/B_t$, $B_t=\hbar c/(2el^2)$. As is seen the changes in
magnetic field dependencies of negative magnetoresistance with
change of inter-layers transition probability reflect the
variation of area distribution functions. Indeed, $\Delta \sigma
(B_z)$ depends on $t$ slightly: maximal change is less than two
times for decreasing $t$ up to zero, whereas $\Delta \sigma
(B_x)$ changes drastically. It decreases about hundred times,
when the value of $t$ decreases from 0.1 to 0.002.

\section{Discussion}
 \label{sec:dis}
Let us compare the calculated area distributions with
experimental data. One can see from  Figs.\ \ref{fig30}a and
\ref{fig40}, that the behaviour of calculated  and  experimental
$W(S_z)$ and $W(S_x)$ dependencies is close qualitatively.  As
mentioned above, $W(S_x)$ depends on inter-layers transition
probability significantly stronger than $W(S_z)$. Therefore we
have estimated the transition probability comparing the
calculated and experimental $W(S_x)$ curves. The most accordance
has been obtained with $t\simeq 0.1$ (see Fig.\ \ref{fig30}a). As
seen from the figure, with this value of $t$ the calculated
$W(S_z)$ dependence describes the experimental data well. Some
quantitative inconsistency, especially for $W(S_x)$, is evident
in Fig.\ \ref{fig30}a. We believe, this is result of crudity of
model  used. In particular, we supposed the identity of both 2D
layers.

Let us turn now to magnetic field dependencies of negative
magnetoresistance. To calculate $\Delta\sigma(B)$, in addition to
the inter-layers transition probability it is necessary to know
the phase breaking length. Using the value of $t=0.1$ estimated
above, we have found that the best agreement between theoretical
and experimental $\Delta \sigma (B_i)$ dependencies is obtained
with $l_\varphi\simeq 3.4$ and $1.4$ $\mu$m for T=1.5 and 4.2 K,
respectively. The $\Delta \sigma (B_z)$ and $\Delta \sigma (B_x)$
dependencies calculated with these $l_\varphi$ values practically
coincide with those measured experimentally (see Fig.\
\ref{fig10}). It should be noted that these values of $l_\varphi$
some differ from those obtained by fitting of the
$\Delta\sigma(B_z)$ curves to the Hikami expression:\cite{hik}
the fit gives $l_\varphi\simeq 4.8$ and $1.7$ $\mu$m for T=1.5
and 4.2 K, respectively. The reason of this difference is that
the Hikami formula was obtained for single 2D layer, and it is
not suitable for analysis of negative magnetoresistance in
coupled double layers structures.

Finally, knowing the values of $t$ and $l_\varphi$ we are able to
compare the calculated and experimental area dependencies of
$\overline{L}(S_x)$ to $\overline{L}(S_z)$ ratio (Fig.\
\ref{fig30}b). It is seen that the experimental ratio is
significantly larger than unity as well as calculated one.
However, the experimental points lie somewhat below than
calculated curve. We suppose that the main reason of such
disagreement is dissimilarity of the layers in structure
investigated.

After this paper has been prepared for publication, the paper by
Raichev and Vasilopoulos on the theory of weak localization in
double quantum wells  is appeared in Condensed Matter e-Print
archive.\cite{raichev} Let us apply this theory to our case.
Using the formulae derived in Ref.\ \onlinecite{raichev}, we have
calculated $\Delta\sigma(B_i)$ dependencies for our structure.
These dependencies are represented in Fig.\ \ref{fig10} by dashed
curves. As is clearly seen, theory developed in Ref.\
\onlinecite{raichev} describes our experimental results only in
low magnetic fields. The reason is that the calculations in Ref.\
\onlinecite{raichev} were carried out in the framework of
diffusion approximation. It means that two conditions are met.
The first condition is $\tau\ll \tau_\varphi$. For the structure
investigated $\tau/\tau_\varphi=0.014-0.035$ for different
temperatures, and this condition may be considered as fulfilled.
According to the second condition, the magnetic field has to be
low enough: $B\ll B_t$ when $\bf B\parallel z$, or $B\ll B_t
l/Z_0$ when $\bf B\parallel x$. In our case $B_t\simeq 0.14$ T,
$l/Z_0\simeq 2.5$ and hence the diffusion approximation is
applicable,  when $B\ll 0.14$ or $0.35$ T depending on the
magnetic field orientation. It is in this range of magnetic field
that the results of Ref.\ \ \onlinecite{raichev} are close to our
experimental data.

Our calculations are valid beyond the diffusion approximation and
therefore they better describe the experimental results in whole
magnetic field range, where weak localization correction to the
conductivity is dominant.

\section{conclusion}
 \label{sec:con}
We have investigated the negative magnetoresistance in double
layer heterostructures for different magnetic field orientations.
The information about statistics of closed paths has been
extracted from the analysis of temperature and magnetic field
dependencies of conductivity. Significant difference in area
distribution functions, $W(S_x)$, $W(S_z)$, and in average
lengths of closed paths, $\overline{L}(S_x)$,
$\overline{L}(S_z)$, has been found. In order to interpret the
experimental results, we have investigated the statistics of
closed paths and negative magnetoresistance using the computer
simulation of the carrier motion with scattering over two 2D
layers. Analysis of experimental and theoretical results
unambiguously shows that in parallel magnetic field the negative
magnetoresistance in double layer structures is determined by
inter-layers transitions.

\subsection*{Acknowledgments}
This work was supported in part by the RFBR through Grant
00-02-16215, and the Program {\it University of Russia} through
Grants 990409 and 990425.

\end{multicols}

\begin{references}
\bibitem{Falko}In fact, in single 2D layer structures the
longitudinal negative magnetoresistance is observed too, but its
magnitude is significantly less than the magnitude of transverse
magnetoresistance. In the structures with one subband filled this
effect arises due to asymmetry of the wave function or scattering
potential with respect to the center of 2D layer.\cite{Falko1}
\end{references}
\end{document}